\documentclass[10pt, journal,letterpaper]{IEEEtran} 
\IEEEoverridecommandlockouts
\usepackage{cite}
\usepackage{amsmath,amssymb,amsfonts}
\usepackage{algorithmic}
\usepackage{graphicx}
\usepackage{textcomp}
\usepackage{url}
\usepackage{hyperref}

\def\BibTeX{{\rm B\kern-.05em{\sc i\kern-.025em b}\kern-.08em
    T\kern-.1667em\lower.7ex\hbox{E}\kern-.125emX}}
\begin{document}

\title{Covert backscatter  communication with directional MIMO}%\\

\author{\Large Roberto Di Candia, Saneea Malik,  Huseyin Yi\u{g}itler, and Riku J\"antti~\IEEEmembership{}
        % <-this % stops a space
\thanks{The authors acknowledge support from Academy of Finland, grant no. 319578. Roberto Di Candia acknowledges support from the
Marie Sk{\l}odowska Curie fellowship number 891517 (MSC-IF
Green-MIQUEC).}% <-this % stops a space
\thanks{The authors are with the Department of Communications and Networking, Aalto University, Espoo 02150, Finland.}
\thanks{Correspondence to: rob.dicandia@gmail.com, riku.jantti@aalto.fi.}}

\maketitle

\begin{abstract}
We study a backscatter communication protocol over a AWGN channel, where a transmitter illuminates a tag with a directional multi-antenna. The tag performs load modulation on the signal while hiding its physical presence from a warden. We show that, if the transmitter-to-tag channel is inaccessible to the warden, then $\Theta(n)$ reliable and covert bits can be transmitted over $n$ channel usages. This overcomes the square-root law for covert communication.
This paper provides the first evidence for practical implementation of covert backscatter communication, with potential applications in 
IoT security.  
\end{abstract}


\section{Introduction}
With the emergence of IoT, the demand of energy consumption rose considerably. IoT is composed of smart technologies and sensors which make up an embedded system for exchanging and connecting other systems or devices, for the exchange of data over the internet~\cite{8862269}. Security risks and challenges also tend to emerge in IoT, as the therein devices are thought to be low-powered and with low memory. These features make standard cryptography tools challenging to apply. Backscatter communication (BC) rises as a promising paradigm for IoT. For instance, ambient BC has enabled connectivity with off the shelf and battery-free devices~\cite{9310744}. BC has emerged as a new technique for transmitting data by modulating the phase and/or amplitudes of an absorbed WiFi ambient signal~\cite{9055221}. It is believed that flexible and scalable deployment of IoT is achievable with ambient BC while reducing cost and power consumption. This promotes significantly the development of this technology for daily-life routines.

Due to the broadcast nature of backscatter systems, their applications in sensitive scenarios is particularly limited by eavesdropping attacks. Encryption techniques have been used recently to limit third-party attacks, but the trade off between power, size and cost is currently not advantageous~\cite{phy}. Problems related to privacy, authentication and confidentiality are handled by the physical layer in wireless systems, by making variations in public key and private key cryptosystem~\cite{springer}.\\
An important solution to the security problem consists in hiding from the potential eavesdropper that the parties are communicating, i.e. {\it covert communication}. This is known to be limited by the square-root law for Additive White Gaussian Noise (AWGN) channels, stating that at most $\Theta(\sqrt{n})$ bits can be transmitted reliably and covertly over $n$ channel usages~\cite{6584948}. Since covert communication requires the coordination of sparse transmission of $\Theta(\sqrt{n})$ signals over $n$ channels, it requires a pre-shared key of $\Theta(\sqrt{n}\log(n))$ bits. Solutions to this problem are based on using covert key expansion protocols~\cite{PhysRevA.99.052329} or uninformed jammers~\cite{7964713}. 
Recently, covert communication with backscatter systems has been investigated for both classical AWGN~\cite{ccbr,9363596,dirty_cons} and quantum bosonic~\cite{PRXQuantum.2.020316} channels. Here a tag (Alice) embeds the information by {\it passively} modulating a signal broadcasted by Bob~\cite{PRXQuantum.2.020316} (mono-static case), or by a third-party transmitter~\cite{ccbr,9363596,dirty_cons} (bi-static case). While the therein analysis show that covertness can be achieved in backscatter systems, no convincing evidence that this is a practical security solution has been reported yet. In fact, standard solutions to achieve sustainable communication fail in BC: Covert key-expansion requires post-processing too complex for a low-memory tag, while jamming can harm the communication of licensed users.

\begin{figure}[t!]
\centering
\includegraphics[width=8.6cm]{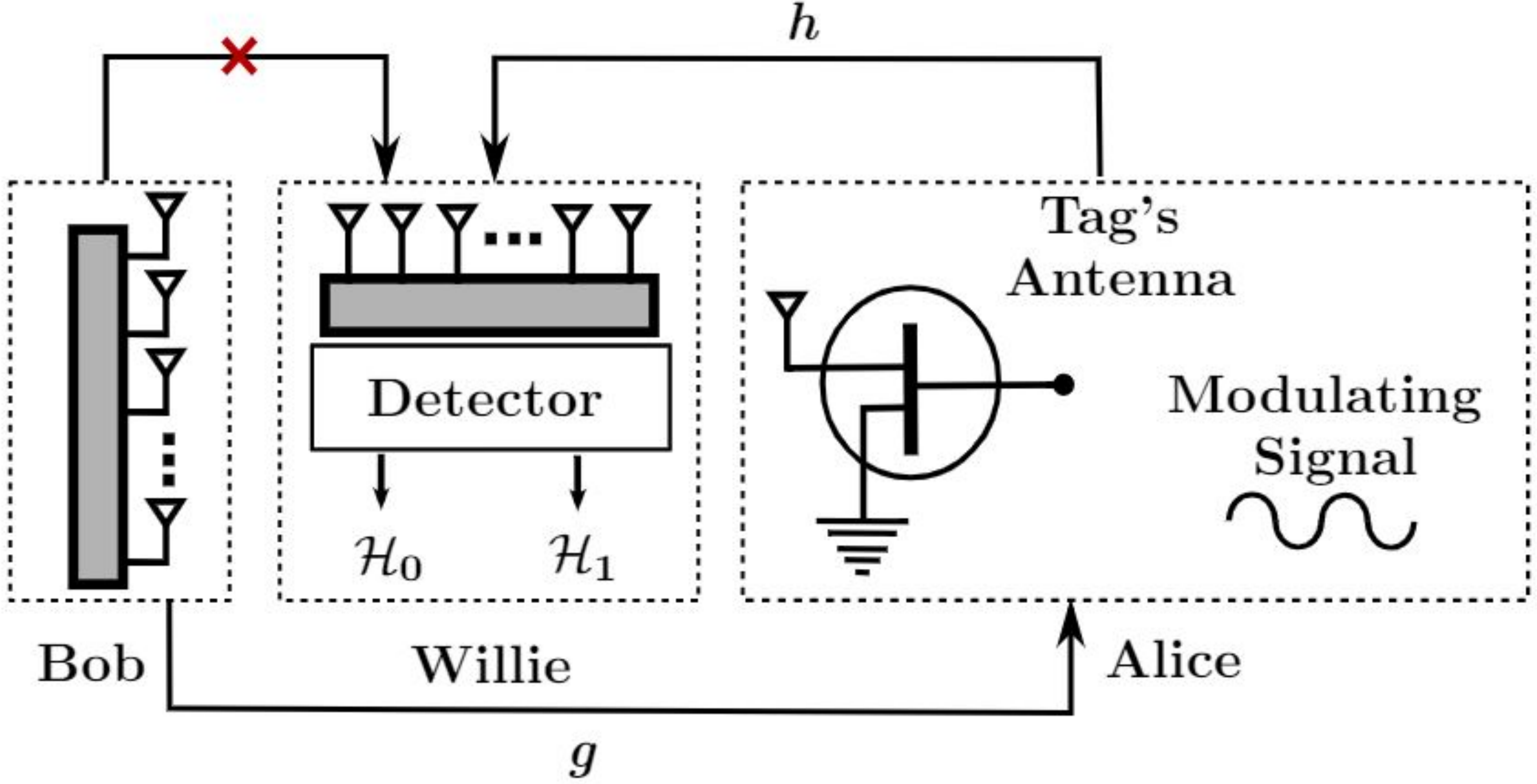}
\caption{Mono-static backscatter system for covert communication. A directional multi-antenna Bob illuminates Alice's antenna. Alice modulates the amplitude, and she then backscatter the signal to Bob. A multi-antenna warden Willie uses receiver beamforming to receive the signal scattered by Alice. It is assumed that there is no direct path between Bob and Willie.
  \label{fig:system}}
\vskip -10pt
\end{figure}

In this paper, we substantially advance towards a practical covert BC implementation, by considering a mono-static backscatter system with a directional multi-antenna as Bob. 
We perform the covertness analysis for a multi-antenna warden Willie with the assumptions (i) Bob-to-Alice channel is secure and (ii) Alice's antenna has a structural mode. Willie is able to catch the radiation from the multiple paths of Alice's scattered signal, see Fig.~\ref{fig:system}. We show that, using a Gaussian illuminating signal, $\Theta(n)$ covert and reliable bits can be transmitted {\it without} a pre-shared secret key. Intuitively, this is possible if one considers Alice's antenna structural mode, that makes Willie's task harder as he/she needs to distinguish between two Gaussian distributions with large variance. It follows that, unlike in \cite{6584948}, communication can be turned-on for {\it all} the $n$ channels, instead of being limited to a $\sqrt{n}$ fraction of modes.
We finally show how to extend the result to the bi-static case.

\section{System model}

We consider a backscatter system, in which Alice communicates with Bob by controlling her antenna's reflection coefficient $\Gamma[t]$, using load modulation.  Without loss of generality, we assume the narrow-band case, meaning that one transmission for each time-step is performed. The extension to the wide-band case is straightforward. Alice's reflection coefficient is given by 
\begin{equation}
    \scriptsize
    \Gamma[t]=\frac{Z[t] - Z_a^*}{Z[t]+Z_a^*},
\end{equation}
where $Z_a\in\mathbb{C}$ denotes the antenna impedance, $Z_a^*$ denotes its complex conjugate,
and $Z[t]$ denotes the controllable load impedance. The detectability of an object can be measured by its complex valued radar cross-section (RCS), that, for a backscatter antenna, can be written as \cite{degen2020complex}
\begin{equation}
    \scriptsize
    \sigma[t]= j \frac{\lambda}{2\sqrt{\pi}}G \left(A_s -\Gamma[t]\right).
\end{equation}
Here, $\lambda$ denotes the wavelength, $G$ is Alice's antenna gain and $ A_s=|A_s|e^{j\psi_s}$ is the structural mode of the antenna. The RCS of a backscatter antenna $|\sigma[t]|^2$ describes the fraction of the illuminating signal power that is scattered back from the antenna. We note that even if Alice's antenna load is matched to the antenna impedance $Z[t]=Z_a^*$ such that antenna absorbs as much power as possible, the device is still reflecting some power due to the structural mode. Let $T_A$ denote Alice's symbol duration. During one symbol duration $kT_A \leq t_k < (k+1) T_A$ a constant load impedance is applied. We assume that the Alice uses efficient semi-passive tag and binary phase shift keying (BPSK) modulation~\cite{5529866} such that $\gamma[kT_A]=A_s x a[k]$ where, $\gamma$ is some constant,  $x$ is an attenuation factor and $a[k]\in\{\pm 1\}$ is the $k^\textrm{th}$ symbol used by Alice.

Bob uses a multiantenna reader having $M_B$ antennas to gather information from Alice who communicates by reflecting Bob's signal.  
The warden Willie uses a multiantenna detector with $M_{W}$ antennas to detect if Alice is communicating. 
We assume that both Bob and Willie do oversampling such that they can collect $L$ samples for each Alice's symbol. Let us define a discrete time slot $(k,l)$ to be the $l^\textrm{th}$ time slot during Alice's $k^\textrm{th}$ information symbol. The duration of a single time slot is $T_A/L$.
Let $d_{B}$ denote the distance between Alice and Bob, $\lambda$ denote wavelength. In line-of-sight (LoS) conditions, the freespace pathloss between Alice and Bob is given by $\eta_{B}=\left(\frac{\lambda}{4\pi d_{B}}\right)^2$. The signal backscattered to Bob from Alice at time slot $(k,l)$ can be written as \cite{Deepak2019}
\begin{equation}
\scriptsize
\tilde{\boldsymbol{y}}_B[k;l] =\sqrt{P_B}(1+x a[k])\boldsymbol{g}\boldsymbol{g}^H\tilde{\boldsymbol{b}}[k;l] + \tilde{\boldsymbol{w}}_B[k;l]
\end{equation}
where $P_B=M_B^2\eta_B^2 P_{tx}$ denotes the total received power when $a[k]=0$, $\boldsymbol{g}\in\mathbb{C}^{M_B}$ denotes the unit length channel vector ($||\boldsymbol{g}||=1$) describing the channel from Bob to Alice, $M_B$ denotes the number of antennas, $\tilde{\boldsymbol{b}}[k;l]\in\mathbb{C}^{M_B}$ is the complex signal transmitted by Bob, and $\tilde{\boldsymbol{w}}_B[k;l]\in\mathcal{C}\left(0, 2\sigma_B^2\boldsymbol{I}\right)$ denotes the  noise at Bob's receiver at time $(k,l)$. We assume that Bob knows the channel and can point a beam towards Alice by choosing $\tilde{\boldsymbol{b}}[k;l]=\boldsymbol{g}b[k;l]$. Bob also performs receiver beamforming to obtain the scalar signal $y_B[k;l]=\boldsymbol{g}^H\tilde{\boldsymbol{y}}_B[k;l]$, that can be written as
\begin{equation}\label{eq:Received_signalB}
\scriptsize
    y_B[k;l]=   \sqrt{P_B} (1+x a[k]) b[k;l] + w_B[k;l].%, \quad l=1,2,...,L
\end{equation}
The signal of interest $a[k]$ is contained only in the real-part, hence Bob only need to consider the real noise $w_B[k;l]\sim\mathcal{N}(0,\sigma_B^2)$ at his detector.

We assume that there is no direct path from Bob to Willie. If Bob has at least the same number of antennas as Willie, and it knows the channel to Willie, it can use nullsteering to achieve this condition. Alternatively, this condition could be satisfied when there is some physical obstacle between the two parties.  In the absence of the direct path, the signal received by Willie at time slot $k$ can be written as \cite{duan2017achievable}
\begin{equation}\label{eq:Received_signalW}
\scriptsize
\tilde{\boldsymbol{y}}_W[k;l] = \sqrt{P_W}(1+x a[k;l])\boldsymbol{h}\boldsymbol{g}^H\tilde{\boldsymbol{b}}[k;l] + \tilde{\boldsymbol{w}}_W[k;l]
\end{equation}
where $P_W=M_WM_B\eta_W\eta_BP_{tx}$, $\eta_W=\left(\frac{\lambda}{4\pi d_{W}}\right)^2$, $d_W$ denotes the distance between Alice and Willie, $\boldsymbol{h}\in\mathbb{C}^{M_W}$ denotes the unit length complex channel from Alice to Willie ($||\boldsymbol{h}||=1$),  and $\tilde{\boldsymbol{w}}_W[k;l]\in\mathcal{C}\left(0, 2\sigma_W^2\boldsymbol{I}\right)$ denotes the  noise at Willie's receiver at time $(k,l)$. 

We assume that Willie knows the complex channel $\boldsymbol{h}$ and can thus point a beam towards Alice. This corresponds to matched filtering and forms sufficient statistics for detection~\cite{Lapidoth2007}.  After considering Bob's transmit beamforming and Willies receiver beamforming, Willie can focus on the real received signal:
\begin{equation}\label{eq:Received_signalW}
\scriptsize
    y_W[k;l]= \sqrt{P_{W}} (1+x a[k]) b[k;l] + w_W[k;l]%, \quad l=1,2,...,L
\end{equation}
where $w_W[k;l]=\boldsymbol{h}^H \tilde{\boldsymbol{w}}[k;l]\sim\mathcal{N}(0,\sigma_w^2)$ is the effective noise seen by Willie. 

We consider two possible illuminating signals for Bob: 1) Constant signal $b[k;l] = 1$ and 2)  Gaussian signal $b[k;l]\sim\mathcal{N}(0,1)$. We note that conditioned on $a[k]$ both $y_W[k;l]$ and $y_B[k;l]$ follow normal distribution. If Bob uses constant illuminating signal, we have 
$y_R[k;l]|a[k]\sim \mathcal{N}(\sqrt{P_{R}} (1+x a[k]),\sigma_R^2)$ for $R\in\{B,W\}$. Similarly, if Bob uses Gaussian illuminating signal, we have $y_R[k;l]|a[k]\sim \mathcal{N}(0,P_R(1+x a[k])^2+\sigma_R^2)$ for $R\in\{B,W\}$.

\section{Performance Analysis}
\subsection{Willie's Detection Error Probability}
Willie uses statistical hypothesis testing based on $n$ consecutive samples between the two hypothesis: communication is happening ($\mathcal{H}_1$), or not ($\mathcal{H}_0$). Here, we assume that Alice transmits one of the $2^m$ equally-likely $m$-bit messages. These are defined by choosing an element from a secret codebook that maps $m$-bits input blocks into $n$-symbol codewords from $\{\pm\}^{n}$, by generating $2^m$ codeword sequences $\mathcal{C}=\{{\bf c}(i)\}_{i=1}^{2^m}$ for message $\{i\}$. Since the codebook is secret, Willie's detection probability is given by averaging over all codebook, with $p({\bf c})=2^{-n}$~\cite{6584948}. This simplifies the analysis, as all symbol transmissions becomes equiprobable independent events. Willie's detection probability distributions for each symbol transmission are $p_W^{}({\bf y}|\mathcal{H}_0)=\prod_{l=1}^Lp_W(y_l|a=0)$ and $p_W^{}({\bf y}|\mathcal{H}_1)=\frac{1}{2^L}\prod_{l=1}^L\left\{p_W(y_l|a=+1)+p_W(y_l|a=-1)\right\}$. 
Over $n$ symbols transmission, we have that
\begin{equation}
   \scriptsize
\mathcal{H}_{0,1}: \quad y_W \sim \prod_{k=1}^n p_W({\bf y}^k|\mathcal{H}_{0,1}),
\end{equation}
%where $[{\bf Y}_{kl}]^{k=1,\dots,n}_{l=1,\dots,L}=({\bf y}^k)_l$.
Assuming equal a-priori probabilities for $\mathcal{H}_0$ and $\mathcal{H}_1$, Willie's average error probability is
\begin{equation}\label{errW}
  \scriptsize
\mathbb{P}_e^{(W)} = \frac{1}{2}-\frac{1}{4}\left\| \prod_{k=1}^n p_W( {\bf y}^k|\mathcal{H}_0)-\prod_{k=1}^n p_W( {\bf y}^k|\mathcal{H}_1)\right\|_{l_1},
\end{equation}
where $\| \cdot\|_{l_1}$ is the $l_1$-norm. One can use the Pinsker's inequality to bound \eqref{errW} as 
\begin{equation}
\scriptsize
    \mathbb{P}_e^{(W)} \geq  \frac{1}{2}-\sqrt{\frac{n}{8}\mathcal{D}_{\rm KL}\left(p_W( {\bf y}|\mathcal{H}_0)||p_W( {\bf y}|\mathcal{H}_1)\right)},
\end{equation}
where $\mathcal{D}_{\rm KL}\left(p( {\bf z})||q({\bf z})\right)= \int_\Omega p({\bf z}) \ln\left(\frac{p({\bf z})}{q({\bf z})}\right)d{\bf z}$ is the Kullback–Leibler (KL) divergence, and we have used its additivity feature for independent distribution. We have that  
\begin{equation}\label{bound}
\scriptsize
\mathcal{D}_{{\rm KL}}\left(p_W({\bf y}|\mathcal{H}_0)||p_W({\bf y}|\mathcal{H}_1)\right)\leq 2\left[ k_m(Y) x^2\right]^2,
\end{equation}
where $m\in\{{\rm c,g}\}$ denotes the constant (c) or Gaussian (g) illuminating signal, and we have introduced the per sample signal-to-noise-ratio $Y=\frac{P_{W}}{\sigma^2_W}$. The bound in \eqref{bound} can be found by performing a Taylor expansion of $\mathcal{D}_{{\rm KL}}$ around $x^2=0$ up to the third order~\cite{6584948}. 

\subsubsection*{Case 1}{\it Constant illuminating signal}. Here, Willie probability distribution is given by
\begin{equation}
  \scriptsize
p_{W, \rm c}({\bf y} | a[k],x) = \prod_{l=1}^L\frac{1}{\sqrt{2\pi}\sigma_W} \exp\left\{-\frac{[y_l-\mu(a[k],x)]^2}{2\sigma_W^2}\right\},   
\end{equation}
where $\mu(a[k],x)= \sqrt{P_{W}}(1+ a[k] x)$. We then obtain 
\begin{equation}
    \scriptsize
k_{\rm c}(Y)=\frac{LY}{2\sqrt{2}}.
\end{equation}

\subsubsection*{Case 2} {\it Gaussian illuminating signal}. Here, the probability distribution of Willie is
\begin{equation}
 \scriptsize
p_{W, \rm g}({\bf y} | a[k],x) = \prod_{l=1}^L \frac{1}{\sqrt{2\pi}\sigma_{W}(a[t],x)} \exp\left\{-\frac{y_l^2}{2\sigma_{W}(a[k],x)^2}\right\},
\end{equation}
where $\sigma_{W}(a[k],x)= P_W(1+xa[k])^2+\sigma_W^2$. We obtain \begin{equation}
\scriptsize
k_{\rm g}(Y) = \frac{LY\left(1+2Y+13Y^2\right)^{1/2}}{2\sqrt{2}\left(1+Y\right)^2}.
\end{equation}
We notice that $k_{\rm c}=O(Y)$ while $k_{\rm g}=O(1)$ for large $Y$. Moreover, $k_{\rm c}/k_{\rm g}\to1$ for $Y\to0$, meaning the Alice's structural mode is relevant only in the Gaussian case. All in all, we have that
\begin{equation}
\scriptsize
\mathbb{P}_e^{(W)} \geq \frac{1}{2}\left(1-k_{\rm m}(Y)x^2\sqrt{n}\right) \geq \frac{1}{2}(1-\varepsilon_W),    
\end{equation}
where $\varepsilon_W$ quantifies the degree of covertness, that can be achieved for $x^2 \leq \left[ k_{\rm m}(Y)\frac{\sqrt{n}}{\epsilon_W}\right]^{-1}\triangleq x^2_{\rm m}$ with ${\rm m}\in \{{\rm c, g}\}$. In the following, we assume  $Y>0$ and $n$ large enough, in order for $x^2_{\rm m}$ to be bounded by one.

\subsection{Alice to BoB Link Performance}
Under hard-decision decoding with bit error probability $p_b$, the capacity for the Alice to Bob link in terms of bits per channel use is given by the capacity of Binary Symmetric Channel (BSC)
\begin{equation} \label{eq:BSC_C}
\scriptsize
    C(p_b) = 1-h_2(p_b) = \frac{1}{2\ln (2)} \left(p_b-\frac{1}{2}\right)^2 + O\left(\left(p_b-\frac{1}{2}\right)^4\right),
\end{equation} 
where $h_2(\delta)=-\delta\log_2(\delta)-(1-\delta)\log_2(1-\delta)$. 
%around $\delta=1/2$. 
Let $ \scriptsize R=\frac{m}{n}$ denote the data rate used by Alice. 
The probability that a code word is erroneously detected is bounded by:
\begin{equation}\label{eq:BLER}
\scriptsize
  \mathbb{P}_{e}^{(B)} \leq e^{-nE_r(R)}  
\end{equation}
where $\scriptsize E_r(R)$ is the Gallagher's random coding error exponent~\cite{thesisRef}. Notice that, in the limit of small signal-to-noise ratio (SNR), the setup corresponds to Class I Very Noisy Channel, where the Gallagher's random coding error can be expressed as~\cite{thesisRef}
\begin{equation} \label{eq:ErrorExponent}
\scriptsize
E_r(R) \simeq \left\{\begin{array}{cc} 
C(p_b)/2-R & 0 \leq \frac{R}{C(p_b)} \leq \frac{1}{4}\\
\left(\sqrt{C(p_b)}-\sqrt{R}\right)^2 & \frac{1}{4}\leq \frac{R}{C(p_b)} \leq 1.
\end{array}\right.
\end{equation}
\subsubsection*{Case 1} {\it Constant illuminating signal}.
Here, Bob's gets the samples $\tilde y_B[k;l]$, and applies the following hard-decision decoding strategy: he decides towards the hypothesis $a[k]=1$ if $\bar y_B[k;l]>\sqrt{P_{B}}b[k;l]$, and $a[k]=-1$ otherwise. 
He uses maximum ratio combining over the $L$ antenna's samples. Let us introduce the SNR 
\begin{equation}\label{SNR}
\scriptsize
    \gamma(x) =\frac{ L P_{B} x^2}{\sigma_B^2}.
\end{equation}
The corresponding bit error probability is $p_b(x_{\rm c}) = Q(\sqrt{\gamma(x_{\rm c})})$, %\cite{proakis1994communication}
where $Q(z)=\frac{1}{\sqrt{2\pi}}\int_z^\infty e^{-\frac{1}{2}t^2}dt$ is the Q-function.
Alice and Bob use a secret codebook to perform error-correction over the $n$ symbol transmissions, as discussed in the previous section.
%since Willie's detection probability is proportional to $P_{tx}x^2$, we have now control of the bit error performance at Bob's receiver and need to cope with the situation using error-correction coding. 
Since $\gamma(x_{\rm c})=\Theta(n^{-1/2})$ does not depend on $P_{B}$, we are in the low SNR region for any $P_{B}$ and sufficiently large $n$. The error probability can be then approximated as $p_b(x_{\rm c})\simeq1/2-\sqrt{\gamma(x_{\rm c})/(2\pi)}$ for sufficiently large $n$.
Using the expansion in Eq.~\eqref{eq:BSC_C}, we obtain the channel capacity under hard-decision decoding: 
\begin{equation}
\scriptsize
    C_{\rm c} \simeq \frac{\gamma(x_{\rm c})}{4\pi\ln (2)} \simeq \frac{1}{\sqrt{2}\pi\ln(2)}\frac{M_B d_W^2}{M_Wd^2_B}\frac{\sigma_W^2}{\sigma_B^2}\frac{\varepsilon_W}{\sqrt{n}}.%\triangleq \kappa_c \frac{\sigma_W^2}{\sigma_B^2}\frac{\varepsilon_W}{\sqrt{n}}.
\end{equation}
We notice that this quantity does not depend on $P_{B}$. The Gallagher's random coding error exponent can be now derived using Eq.~\eqref{eq:BLER}.
In this case, $m=\Theta(\sqrt{n})$ covert bits can be reliably transmitted for large-enough $n$ channel usages.

\subsubsection*{Case 2} {\it Gaussian illuminating signal}.
Here, Bob's gets the samples $\{\tilde y_B[k;l], l=1,2,...,L\}$, and uses matched filtering to obtain $\sum_{l=1}^L b[k,l]y_B[k;l]$. The instantaneous SNR for the symbol $k$ is given by
\begin{equation}
\scriptsize
    \gamma_{\rm g}[k](x_{g}) =\frac{ \sum_{l=1}^L b^2[k,l] P_{B} x_{g}^2}{\sigma_B^2}.
\end{equation}
The corresponding bit error probability is $p_b[k] = Q(\sqrt{\gamma_{\rm g}[k](x_{g})})$.
The situation resembles the one of a fading channel with receiver side information. Therefore, the covert capacity is $C_{\rm g}=\mathbb{E} \left\{C(p_b[k](x_{\rm g}))\right\}$~\cite{wang_2020}. Let us first consider the received $P_{B}$ is fixed as $n$ increases. In this case, the domain of $\gamma_{\rm g}[t](x_{\rm g})$ is essentially small for $n$ large enough. We can use the expansion in Eq.~\eqref{eq:BSC_C} and the fact that $Q(z)\simeq \frac{1}{2}-\frac{z}{\sqrt{2\pi}}$ for $z\ll1$ to obtain
\begin{equation}
\scriptsize
    C_{\rm g}\simeq \frac{\gamma(x_{\rm g})}{4\pi\ln(2)}, 
\end{equation}
where $\gamma(x)$ is defined in Eq.~\eqref{SNR}. \\
Unlike the constant illuminating case, here we have that $\gamma(x_{\rm g})=\Theta\left(\frac{P_{B}}{\sqrt{n}}\right)$. This means that we can choose $P_{B}$ such as to have an arbitrarily large SNR and thus arbitrary small bit error probability $\varepsilon_B$. That is, we choose $P_{B}$ such that $\mathbb{E}\{Q(\sqrt{\gamma_{\rm g}[k](x_{g})})\}=\varepsilon_B$ leading to capacity $C\geq 1-h_2(\varepsilon_B)$. In this case, $m=\Theta(n)$ covert bits can be reliably transmitted for large-enough $n$ channel usages.

{\begin{figure}[t!]
\centering
\includegraphics[width=8.7cm]{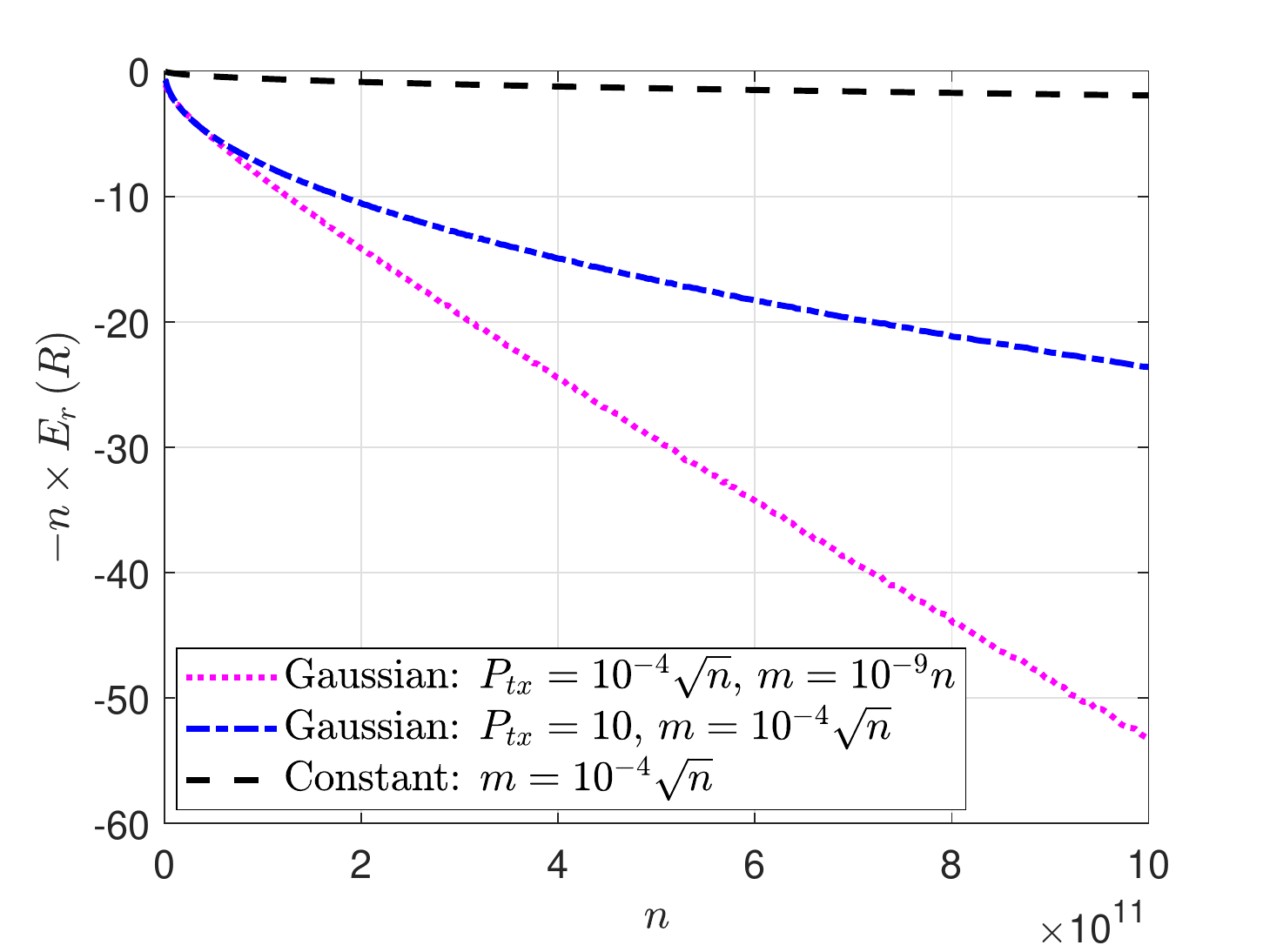}
\caption{Bound on Bob's decoding error probability exponent for the transmission of $m$ symbols, as in Eq.~\eqref{eq:BLER}. The plot has been drawn with exact numerical calculations of the capacity in the various cases. The values of parameters are fixed as $\sigma_{B,W}=1$, $\varepsilon_{W}= 10^{-4}$, $M_{B,W}=100$, $d_{B,W}=1$ $L=10$. The Gaussian case outperforms the constant case for large enough $n$. Moreover, in the Gaussian case, scaling the transmitting power $P_{tx}$ with $\sqrt{n}$ allows to transmit reliably $m=\Theta(n)$ bits.} \label{fig2:plot}
\vskip -10pt
\end{figure}}
\section*{Conclusion and Outlook}
In this paper, we have investigated the impact of the structural mode of Alice's antenna for covert communication. We have considered a mono-static backscatter communication system, where a multi-antenna Bob illuminates Alice's antenna in two scenarios: constant or Gaussian illuminating signal. We have shown that, if the link Bob-to-Eve is secure, then Alice and Bob can communicate $\Theta(n)$ covert bits for large $n$ without need of a pre-share secret. This limit is reached with Gaussian illuminating signals in the large power limit. 

The mono-static scenario can be extended to the bi-static case as follow.
It is easy to see that in the constant illuminating case the same results hold, with no-more assumptions than the one in the mono-static case, i.e. a secret codebook between Alice and Bob. In the Gaussian illuminating case, instead, the transmitter needs to share with Bob the values of the samples $\tilde b[k;l]$, which requires infinite bits for any $n$. An alternative consists in using correlated Gaussian noise~\cite{7921552} at the transmission level, where the signal is sent to Alice and the idler to Bob, in a secure way. The situation resembles the one where Alice and Bob securely share entanglement (quantum correlations) to overcome the square-root law~\cite{9178961}. Here, our results contribute in the following: If the structural mode of Alice's antenna is taken into account, then classical correlations are enough to overcome the square-root law.

\vspace{12pt}

\end{document}